# Molecular Level-Crossing and the Geometric Phase Effect from the Optical Hanle Perspective

R. Glenn[1] and M. Dantus,[1,2]

[1]Department of Chemistry, Michigan State University, East Lansing, Michigan 48824, USA
e-mail address: rglenn@msu.edu
[2]Department of Physics and Astronomy, Michigan State University, East Lansing, Michigan 48824, USA

Level-crossing spectroscopy involves lifting the degeneracy of an excited state and using the interference of two nearly degenerate levels to measure the excited state lifetime. Here we use the idea of interference between different pathways to study the momentum-dependent wave packet lifetime due an excited state level-crossing (conical intersection) in a molecule. Changes in population from the wave packet propagation are reflected in the detected fluorescence. We use a chirped pulse to control the wave packet momentum. Changing the chirp rate affects the transition to the lower state through the conical intersection. It also affects the interference of different pathways in the upper electronic state, due to the geometric phase acquired. Increasing the chirp rate decreases the coherence of the wave packet in the upper electronic state. This suggests that there is a finite momentum dependent lifetime of the wave packet through the level-crossing as function of chirp. We dub this lifetime the wave packet momentum lifetime.

## I. INTRODUCTION

Level-crossing occurs when two levels approach each other and intersect or avoid each other and is a fundamental aspect in quantum physics and chemistry. Level-crossing can be induced by a magnetic field (Zeeman splitting) or an optical field (Stark effect). The external field lifts the degeneracy of the level and the level spacing induced is less than the natural linewidth. This type of level-crossing has been well studied [1]. Larger level spacing, such as an electronic level-crossing, occurs in molecules due to a *cis-* to *trans-* isomerization [2-4]. How the wave packet propagates due to the electronic level-crossing (conical intersection), is still being understood. Here we describe how a chirped pulse affects the wave packet formation and how the chirp rate can control wave packet interference and transport associated with a conical intersection.

The Hanle effect [5] has been used to study level-crossing. It has been observed in atoms [6-8], molecules [9, 10], and in experiments on the optical orientation of spins of electrons in semiconductors [11, 12]. The core of the Hanle effect is that an external magnetic field allows one to control the separation between two crossing energy levels and allows one to control the florescence interference. Figure 1a illustrates the crossing between two split excited state levels $a$, $b$. The coherence or phase relation between the levels alters the coherence in the fluorescence. For example, for well-separated levels (electronic levels), independent emission from the levels occurs and the fluorescence signal is proportional to the sum of the squares of amplitudes $A_a^2 + A_b^2$, where $A_a$ ($A_b$) is the amplitude from the state $a$ ($b$). When both levels are nearly degenerate and excited simultaneously, the fluorescence is proportional to $(A_a + A_b)^2$. Resonance occurs when the magnetic field is zero and increasing the magnetic field strength partially destroys the coherence. Thus, the absolute radiative lifetime can be measured by detecting the fluorescence as a function magnetic field strength, without any knowledge of the density of the emitters [13].

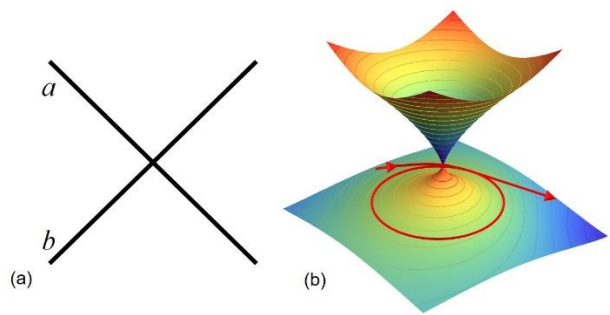

FIG 1. (a) Level-crossing due to the lifting of the degeneracy. (b) Level-crossing between electronic states in molecule (conical intersection). The wave packet propagation acquires a geometric phase when encircling the conical intersection (red line).

Electronic level-crossings (conical intersections) occur in nearly all molecules and are responsible for many excited-state chemical processes [2-4, 14-16]. The isomerization process can be described in the Born-Oppenheimer approximation, where the nuclear motion defines the potential energy surfaces corresponding to different electronic states that intersect at the conical intersection point, in the nuclear coordinate, with the geometric topology of a double cone (Fig. 1b). The two potential energy surface topologies and their interaction is influenced by the molecular structure [17, 18], medium [19-21], and the symmetry of the two states [2, 14, 22]. It also depends upon whether the molecule isomerizes in the ground or in the excited state [23]. The presence of a conical intersection is recognized by the rapid non-radiative transfer of energy from the upper to the lower state and the geometric phase effect [16, 24-32]. The rapid decay of the electronic excited state and lack of fluorescence is a commonly known description that may not apply to every molecule. For example, upon two-photon excitation to the $S_2$ state more fluorescence was observed from $S_2$ than from $S_1$. This is because of a rapid decay of the $S_1$ to $S_0$ (ground state) as a result of a conical intersection [33]. Examples of highly fluorescent molecules with conical intersections do exist [34]. Transient absorption measurements show indications of wave packet bifurcation in the potential energy surface due the conical intersection [34].

The main signature of the geometric phase effect is the destructive interference in population transport due to a conical intersection. The interference quenches the fluorescence [16, 24-29]. The geometric phase is sensitive to the wave packet's angular momentum [35, 36], which can be controlled by tuning the spectral phase of the laser pulse [37]. Theoretically, angular momentum is added to a wave packet by multiplying it by $e^{il\theta}$, where $\theta$ is the phase, and $l$ is the angular momentum [36]. A wave packet with vanishing angular momentum encircling the conical intersection, Fig. 1b, will spread more than a wave packet with high angular momentum and thus decohere faster and decrease the fluorescence yeild [36]. Determination of the optimal pulse to control the wave packet's phase (momentum) and thus its transport through the conical intersection has been done by means of genetic algorithms [38-40] and numerical simulations [41, 42]. Alternatively, shaped pulses can change the wave packet profile by multiple interactions with the same field while in the excited state or by multiple transitions to the excited state [35, 43, 44]. Controlling the wave packet profile via shaped pulses and its transport through a conical intersection for molecules in solution has been demonstrated [38-40, 45].

As stated earlier, the molecular Hanle effect involves changing the magnetic field strength to affect the interference of light waves emitted from two excited states in the course of measuring the fluorescence. Here we propose that by analogy with the Hanle effect, varying a parameter of the laser, such as its chirp rate, the wave packet interference associated with electronic level-crossings (conical intersection) can be controlled. The resulting control of the wave packet can be detected in the fluorescence.

We consider a fluorescent molecule with three electronic states $S_0$, $S_1$, and $S_2$. The $S_2$ state is directly excited by a chirped laser pulse and the transition to the $S_1$ state occurs non-adiabatically via a conical intersection. The creation of the vibrational coherences in the excited state $S_2$ is controlled by the chirp rate. For a chirped pulse, the carrier frequency varies linearly in time $\omega(t) = \omega_\alpha t$, where $\omega_\alpha$ is the chirp frequency, directly affecting the phase of the wave packet as a function of time [46]. Using Floquet theory, one obtains a solution to the time-dependent Schrodinger equation with a chirped pulse in the form of a linear combination of eigenstates. The eigenstates evolve periodically in time with frequency $\omega_\alpha$. The wave packet consists of a linear combination of stationary vibrational states with energy proportional to $e^{i\varepsilon_n\theta}$, where $\theta = \omega_\alpha t$, and $\varepsilon_n$ is the associated vibrational eigenvalue [47, 48]. Tuning the chirp parameter allows us to control the angular momentum of the wave packet and thus its deconstructive interference and transport through a conical intersection. Similar to the Hanle effect, we expect that the fluorescence yield as a function of chirp will have a peak at zero chirp (transform limited). Increasing the chirp rate increases the angular momentum and this will decrease (increase) the fluorescene yeild due to wave packet deconstructive (constructive) interference. This suggests that a lifetime associated with the wave packet interference due to the conical intersection (at the center frequency of the pulse) can be deduced from the chirped rate width, which we refer to as the wave packet momentum lifetime.

This paper is organized as follows. In Sect. II we derive and anylze the expressions for the fluorescence from the states $S_1$ and $S_2$. In Sect. III we simulate the fluorescence using the theory. Concluding remarks are presented in Sect. IV.

## II. THEORY

The theoretical model aims to describe the fluorescence from a solvated molecule with a ground state $S_0$ and two electronic states $S_1$ and $S_2$. The $S_1$ state is located well outside the laser spectrum. The states $S_1$ and $S_2$ are coupled by a conical intersection. Many of the theoretical advances describing molecular isomerization have been done with quantum chemistry [2, 49-52] and semiclassical [53-55] calculations. Quantum chemistry simulations of the excited electronic state isomerization and the wave packet propagation of a solvated molecule is still a difficult task to complete. Inclusion of the solvent makes the wave packet relax faster in the excited state and it follows the curvature of the potential surface until it reaches the bottom of the well.

Also, the solvent polarity changes the relative energies of the excited state potential energy surfaces and effects the conical intersection point [56, 57], which is reflected in the fluorescence [58]. When describing the fluorescence for weak electric fields, time-dependent perturbation theory can be used [59]; however, the calculated signal is not sensitive to the phase of the pulse. We gave consideration to using the density matrix for describing the processes occurring in our system. The density matrix cannot account for the interference due the geometric phase effect, since it vanishes in the diagonal elements [60]. Here we use an analytical theory to incorporate the geometric phase acquired and the wave packet interference, as demonstrate below.

The theoretical description is organized as follows. First, we derive the population transfer from the ground state to the excited state $S_2$. This initial population $S_2$ can non-adiabatically transition through the conical intersection to $S_1$ or remain in $S_2$. Next, we describe the pathways contributing to the $S_2$ fluorescence and the deconstructive interference of these pathways due the geometric phase acquired. Then the non-adiabatic transition probability to the $S_1$ state is discussed. Finally, the expressions for the population of states $S_1$ and $S_2$ are presented.

### A. Population transfer to the $S_2$ state.

Consider a two-level system, $S_0$ and $S_2$, when excited by a single chirped pulse $E(\omega) = \tilde{E}_0 \tilde{A}(\omega) \exp[i\frac{\alpha}{2}(\omega - \omega_0)^2]$ where $\tilde{E}_0$ is the amplitude of the pulse, $\tilde{A}(\omega)$ is a Gaussian envelope, $\alpha$ is the chirp rate in $fs^2$ and $\omega_0$ is the center frequency of the pulse. The pulse in the time-domain is found by using the Fourier transform and is given as

$$E(t;\alpha) = E_0 A(t;\alpha) e^{-i\omega(t;\alpha)t - i\omega_0 t} \quad (1)$$

where $A(t;\alpha) = e^{-t^2/2\tau^2}$ is the Gaussian envelope. In the time-domain, the spectral chirp increases the pulse width as $\tau = \tau_0\sqrt{1 + \alpha^2 \tau_0^{-4}}$. The frequency changes as $\omega(t) = \omega_\alpha t$, where $\omega_\alpha = \frac{\alpha}{2}(\tau_0^4 + \alpha^2)^{-1}$ and the peak intensity decreases as $I = I_0(1 + \alpha^2 \tau_0^{-4})^{-1/2}$. For a chirped pulse there are regions where the pulse intensity changes slowly or rapidly. This can be seen in Fig. 2a, where $\alpha < 1000\,fs^2$ the amplitude changes rapidly while for $\alpha > 1000\,fs^2$ the amplitude changes slowly. This indicates that there are two different regions. One region is where the chirp rate will have a physical effect on population transfer to the excited state. The other region is in the limit of larger chirp rates, where the intensity changes very slowly, and the chirp rate will have a limiting asymptotic effect on the population transfer. The maximum value of the chirp rate in the time-domain can be found from the detuning frequency $\omega(t)$ and is given as $\alpha_{\max} = \tau_0^2$. This does not define the transition point between the two regions, it simply gives an indication where the chirp rate is greatest.

The time evolution of the system is given by the Schrodinger equation $\frac{d}{dt}\Psi(t) = -iH(t)\Psi(t)$, where $\Psi(t) = \psi_{S_0}(t) + \psi_{S_2}(t)$. Here we have set $\hbar$ equal to one. The Hamiltonian can be written as

$$H = \begin{pmatrix} 0 & \mu A(t;\alpha) \\ \mu A(t;\alpha) & \Delta(t;\alpha) - i\Gamma \end{pmatrix} \quad (2)$$

where $\Delta(t;\alpha) = (\omega_{S_2} - \omega_{S_0}) - \omega_0 - \omega(t)$ is the detuning from $S_0$ and $S_2$, $\mu$ is the transition dipole moment, and $\Gamma$ is the decay rate. We performed a substitution for $\psi_{S_2}(t)$ to bring the linear frequency shift from the pulse eq 1 into $\Delta(t;\alpha)$, so that the pulse is a Gaussian $A(t;\alpha)$.

For $|i\Delta(t;\alpha) + \Gamma|$ large compared to the width of the Gaussian envelope $\tau^{-1}$ an analytical solution can be derived by setting $\frac{d}{dt}\psi_{S2}(t)$ equal to zero on the left hand side of Schrodinger's equation and solving the coupled equations [61, 62]. The $S_0$ population can be found as

$$P_{S_0} = C \exp\left[-\frac{\Gamma}{2}\int_{-\infty}^{\infty} \frac{\mu^2 A^2(t;\alpha)}{\Delta^2(t;\alpha) + \Gamma^2} dt\right] \quad (3)$$

where $C$ is a constant. From eq 3, we see that the population transfer to the excited state $(1 - P_{S_0})$ is insensitive to $\pm \Delta(t;\alpha)$. The population of the excited state $(1 - P_{S_0})$ is shown in Fig. 2b. The population transfer is minimum for a transform limited pulse ($\alpha = 0$) and increases as the chirp rate increases, whether it's sign is positive or negative, until it reaches an asymptotic value of one. The absorption-like peak shown in Fig 2b becomes broader when the intensity of the pulse decreases. Thus decreasing the intensity of the pulse decreases the population of the excited state or the fluorescence yield. By inserting a factor in the argument of the exponential, eq 3, we can simulate any decrease of the population reaching the bottom of the potential energy surface, which we will do later in the final expressions.

Wave packet propagation in a steep potential energy surface can lead to an asymmetric behavior with positive vs. negative chirp rates [63, 64]. This is because for a negatively chirped pulse, the low frequencies trail the high. Initially the pulse excites higher energy modes and they propagate to lower energy modes, the interaction between the already occupied lower energy modes and exciting low frequency photons causes stimulated emission and quenches the fluorescence rate. The steeper the potential energy curvature, the greater the excited state depletion. A positively chirped pulse has the opposite time ordering, so this does not occur. This is why more fluorescence can observed for positively chirped pulses than negative. The steepness of the potential energy surface can influence the amount of population reaching the conical intersection, for negatively chirped pulses, depending on the particular system. Here we assume that there is no asymmetric population depletion due the chirp rate prior to the wave packet reaching the conical intersection.

Here we consider photon energies below the ionization threshold, in the visible and UV range. For photon energies in the x-ray range, theoretical studies revealed that a pump and chirped probe pulse can be used to measure electron dynamics through wave packet motion and interference between two electronic states in a diatomic molecule [65].

Initially, the population is created in the $S_2$ excited state $P_{S_2} = 1 - P_{S_0}$. However, after a few hundred picoseconds the molecule fluoresces from both $S_1$ and $S_2$. This implies that the fluorescence can be found by replacing $P_{S_2} \to P_{S_1} + P_{S_2}$. Now the system is described by the relation $P_{S_1} + P_{S_2} = 1 - P_{S_0}$. This suggests that both $P_{S_1}$ and $P_{S_2}$ have a form proportional to $1 - P_{S_0}$ with the ground state population given by eq 3. Similar forms of eq 3, $\propto (1 - P_{S_0})$, have been used to describe population transfer in avoided crossings with an intense continuous wave chirped laser [66], adiabatic rapid passage population transfer in a two [67-69] and four-level systems [67]. It has been studied for a pulsed excitation of a two-level system [61]. A mathematical formalism similar to what we used in derivation of eq 3 has been used to describe the population transfer for a two-level system when the nuclear motion changes the electronic energy position linearly (linear change in frequency) [70].

The shape of the peak shown in Fig. 2b has been observed experimentally for Rb atoms [68]. Increasing the chirp rate increases the amount of initial excitation and increases isomerization yield [37], and has been verified experimentally [35, 38, 40]. Therefore, the fluorescence yield from $S_1$ and $S_2$ should track with the form of eq. 3. In the next section, we will describe how wave packet interference due to the geometric phase acquired along different pathways changes the chirp dependence from an absorption-like shape (Fig. 2b) to an emission-like shape.

**B. Geometric phase and wave packet mixing**

The wave packet pathways to reach the minimum of the $S_2$ potential energy surface are shown in Fig. 3. The black dashed lines show one pathway directly reaching the minimum of the potential energy surface $S_2$ or another encircling the conical intersection and then reaching the minimum. When the wave packet encircles the conical intersection it acquires a geometric phase of $\pi$ (sign change) [16, 24-29]. The Appendix describes how the geometric phase can be accounted for by a pseudo magnetic field, i.e., a vector potential inserted into the effective nuclear Hamiltonian [71]. When the wave packet propagates around the conical intersection, it will have a preferred direction depending upon the sign of its angular momentum.

The geometric phase is observed when the two pathways mix at the minimum of the potential energy surface [60, 72-75]. The mixing of the two pathways will depend upon the wave packet momentum. For example, wave packet components with low momentum will relax directly to the minimum of the potential energy surface, since they do not have enough momentum to make it over the barrier to the conical intersection. Wave packet components with higher momentum will make it over the barrier to $S_1$ or encircle the conical intersection and stay in $S_2$. In the simplest description, this can be described as a bifurcation of the wave packet in the excited state potential energy surface $S_2$

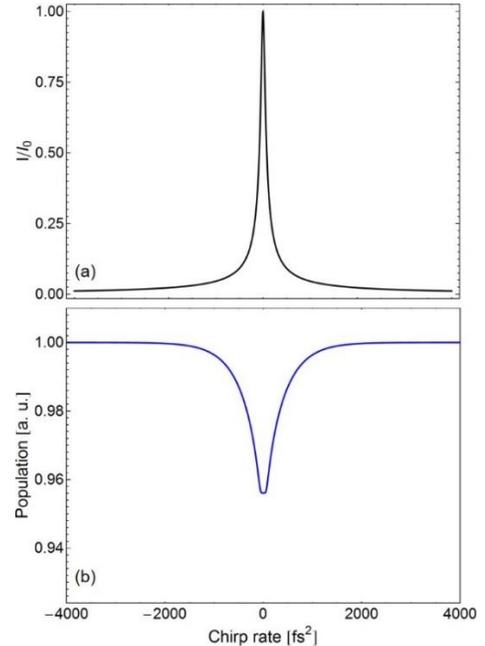

FIG 2. (a) The peak intensity of a chirped Gaussian pulse with width 16 fs. (b) The initial population of the excited states $1 - P_{S_0}$ as a function of chirp rate calculated using 16 fs Gaussian pulse with $\Gamma = 3$ps, $\omega_0 = 545$nm, $\omega_{S_2} = 581$nm, $C = 0.044$, field strength $E_0 = 0.044$.

$\left| \psi_{S_2} \right\rangle \left\langle \psi_{S_2} \right| = \frac{1}{\sqrt{2}} \left( \left| \psi_{S_2}^A \right\rangle \left\langle \psi_{S_2}^A \right| + e^{iC(\gamma)} \left| \psi_{S_2}^B \right\rangle \left\langle \psi_{S_2}^B \right| \right)$, where one

wave packet encounters the conical intersection and acquires a geometric phase $|\psi_{S_2}^B\rangle\langle\psi_{S_2}^B|$ and the other does not $|\psi_{S_2}^A\rangle\langle\psi_{S_2}^A|$. The mixing of the two wave packets creates a population proportional to [60]

$$P_{S_2} = 1 + \left|\langle\psi_{S_2}^A|\psi_{S_2}^B\rangle\right|\cos\left[C(\gamma) - \arg\langle\psi_{S_2}^A|\psi_{S_2}^B\rangle\right] \quad (4)$$

where $C(\gamma)$ is the geometric phase acquired after propagating around the conical intersection. See the Appendix for the derivation of $C(\gamma)$. Here we assume only one conical intersection $C(\gamma) = -\pi$ and that the argument in the cosine is zero. The mixing of the wave packet components from these two pathways creates destructive interference (eq 4) reducing the fluorescence rate of $S_2$. If the wave packet encircles the conical intersection an even number of times the sign change cancels. However, this is not expected, since many conical intersections involve a barrier that the wave packet must overcome [36].

There are a couple reasons why the pathways involve the minimum of $S_2$. First, the transition probability to the $S_1$ state is maximum at the conical intersection point and decreases as $1/r$, where $r$ is the distance between the crossing point and the transition point on the potential energy surface [76]. For a molecule in the condensed phase, wave packet relaxation occurs faster and follows the potential energy surface [77].

An important requirement for observing the geometric phase effect is to have an excitation pulse with sufficient energy and bandwidth. Then the initial wave packet will have high enough momentum and a broad range of momenta to experience multiple pathways in the potential energy surface with one of them circling the conical intersection. and go to the $S_1$ state or encircle the conical intersection returning to the $S_2$ minimum and create destructive interference, quenching the fluorescence from $S_2$.

Chirping the pulse increases the population in the excited state and increases the momentum of the wave packet [37, 46, 47]. This would suggest that the wave packet has more momentum to overcome the barrier and encircle the conical intersection (black-dashed line in Fig. 3) or transition non-radiatively to the $S_1$ state (red line in Fig. 3). Thus by increasing the chirp rate, we increase the amount of wave packet components encircling the conical intersection and thus decrease the $S_2$ fluorescence. At the same time we increase the amount of wave packet components transitioning to $S_1$, thus increasing the $S_1$ fluorescence. The sign of the chirp, while it may affect the direction of the wave packet motion around the conical intersection, does not change the effect of chirp rate on the final populations in this system.

### C. Non-adiabatic transfer to the $S_1$ state.

The transition probability from $S_2$ to $S_1$, $\wp_{S1}$, via the conical intersection can be calculated using semi-classical nonadiabatic Landau-Zener theory (Zhu-Nakamura theory) [78, 79]. The transition probability to stay in the $S_2$ state is then given by $\wp_{S2} = 1 - \wp_{S1}$. Landau-Zener does not include the effects of the geometric phase, however, it is widely used with simulating systems that have the geometric phase effect [76, 78-81]. The theory can incorporate either a one-dimensional or two-dimensional curvature of the potential energy surfaces and their coupling into the calculation of the transition probability.

### D. Final expressions for Fluorescence.

The net population transfer from the $S_0$ to $S_2$ is given as $1 - P_{S_0}$ using eq 3. Equation 4 describes the wave packet bifurcation and destructive interference in the excited $S_2$ state, which is general and does not depend upon the chirp rate. Now, we will combine the population transfer, eq 3, and the wave packet deconstructive interference, eq 4 to determine the final expression for the fluorescence from $S_2$.

As the chirp rate increases beyond its' maximum, $\alpha_{max} = \tau_0^2$, the excitation pulse become longer and its' intensity changes slowly. In the large chirp limit, the effect of the chirp rate on the fluorescence of $S_2$ reaches an asymptotic value and can be described by its effective probability to remain in that state $\wp_{S_2}$. See Section IIC. The effect of the chirp rate on the population transfer from the ground to the excited state $S_2$ is described by the second term in eq 3. However, the mixing of the two pathways reaching the minimum of $S_2$ (Fig. 3 and eq 4) creates destructive interference, as the chirp rate is increases. The deconstructive interference can be incorporated by changing the sign of the exponential in eq 3. The population of $S_2$ can be written as

$$P_{S_2} = \wp_{S_2} + C_{S_2} \exp\left[-\beta_{S2} \frac{\Gamma}{2} \int_{-\infty}^{\infty} \frac{\mu^2 A^2(t)}{\Delta^2(t)+\Gamma^2} dt \right] \quad (5)$$

where, $C_{S2}$ is a constant and $\beta_{S2}$ is a constant that is less than one. The term $\beta_{S2}$ was inserted to describe the distribution of initial population $(1-P_{S_0})$ between the two states $S_1$ and $S_2$ as a function of chirp. Decreasing $\beta_{S2}$ will widen the width of the exponential in eq 5 and decrease the population measured in the fluorescence. Equation 5 suggests that the fluorescence is maximum when using a transform limited pulse and decreases as the chirp rate increases until it reaches its asymptotic value.

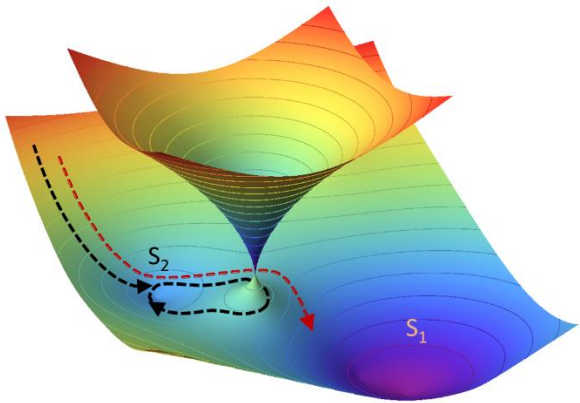

FIG 3. Illustration of the excited state wave packet pathways. $S_1$ and $S_2$ are connected by a conical intersection. The wave packet is initially excited in $S_2$. Low-momentum wave packet components propagate directly to the bottom of $S_2$. Higher momentum wave packet components either overcome the barrier

The fluorescence from $S_1$ can be expressed in the same manner, however, increasing the chirp rate increases the transition to $S_1$ [34, 37]. The population of $S_1$ can be written as

$$P_{S_1} = \wp_{S_1} - C_{S_1} \exp\left[-\beta_{S1} \frac{\Gamma}{2} \int_{-\infty}^{\infty} \frac{\mu^2 A^2(t)}{\Delta^2(t)+\Gamma^2} dt \right] \quad (6)$$

where $C_{S1}$ and $\beta_{S1}$ are constants. Equations 5 and 6 represent the population of the two states $S_1$ and $S_2$ coupled by a conical intersection. It is important to note that a similar form of eq 6 was derived for intense chirped pulses in the dressed state representation [66]. They obtained similar expressions as eq 6 for the population transfer between crossing energy levels. Here we consider a Gaussian shaped pulse and incorporate the geometric phase effect into the population transport with an electronic level-crossing due to a conical intersection.

## III. SIMULATION

Equations 5 and 6 are used to simulate the fluorescence probabilities of $S_1$ and $S_2$ individually. The transition probabilities, $\wp_{S1}$ and $\wp_{S2}$ in eqs 5 and 6 are calculated using Zhu-Nakamura theory mentioned in Sect. IIC. In this model system, we use two Morse potentials that have a minimum corresponding to the fluorescence frequency, shown in Fig. 4. The transition region was considered the region between the minimum of the $S_2$ state and the minimum of the upper adiabatic state, see the black arrow in Fig. 4. The coupling between the states was considered linear $V_{S_1S_2}(R) = cR$, where $c = 0.0008$ is the coupling between states in Hartree units [76]. The transition probability to the $S_2(S_1)$ state in the large chirp-rate limit was calculated as $\wp_{S2} = 0.75$ ($\wp_{S1} = 0.25$). This reflects what is commonly expected for a conical intersection, which is that the majority of population transfer occurs non-adiabatically to the lower state [33].

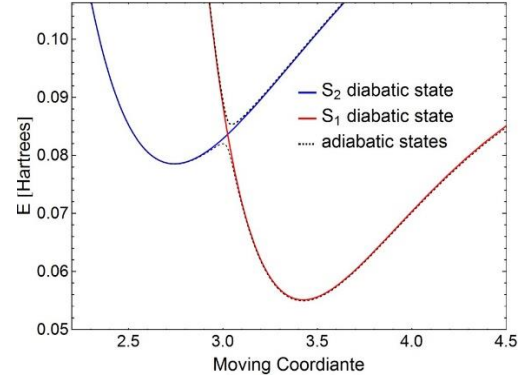

FIG 4. The non-adiabatic potential energy surfaces used to calculate the non-adiabatic transition between $S_1$ and $S_2$.

To simulate the distribution of population among the two states and reproduce the relation $P_{S_1} + P_{S_2} = 1 - P_{S_0}$ the parameters where chosen as $\beta_{S2} = 0.76$, $\beta_{S1} = 0.72$, $C_{S1} = 0.20$, and $C_{S2} = 0.25$. The initial population $1-P_{S_0}$ (blue-dashed line) is shown in Fig 5b with $P_{S_1} + P_{S_2}$ (red-solid line) and this shows that the model system satisfies the relation $P_{S_1} + P_{S_2} = 1 - P_{S_0}$. The broadening of the feature from the initial population Fig. 5b to the final population Fig. 5a reflects the distribution of the population measured in the individual $S_1$ and $S_2$ states as a function of chirping the pulse. This suggests that the linewidth of the peaks seen Fig. 5a are sensitive to the chirp rate or the wave packet momentum, therefore, we identify the width as a momentum-dependent lifetime. Measurement of the lifetime of $S_1$ and $S_2$ in Fig 5a can be easily done, since the chirp rate can be related to the full width half max (fwhm) of the pulse. The fwhm never goes to zero, so we need to subtract the fwhm at zero chirp. In this case, the fwhm of the pulse at the positive chirp fwhm point is subtracted from 16.0 fs, and divided by 2. Division by 2 is necessary because we want to measure half of the fwhm. The same is done for the fwhm point for the negative chirp value. This gives a lifetime of

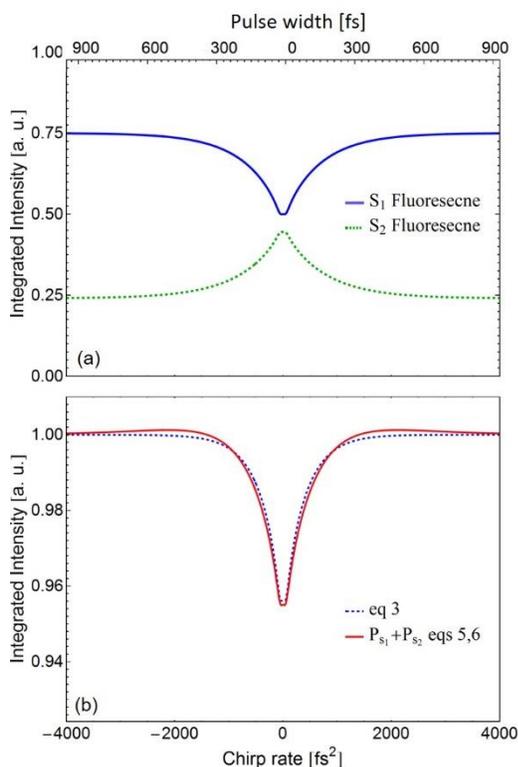

FIG 5. (a) The simulated fluorescence from $S_1$ and $S_2$ using eqs 5 and 6. See the text for the parameters. (b) The initial population ($1-P_{S_0}$), the same as shown in Fig. 2b (blue-dashed), is compared to $P_{S_1}+P_{S_2}$ (red-solid) from (a).

150 fs for the $S_1$ state and 180 fs for the $S_2$ state. Since population transport is sensitive to the intensity of the pulse, we expect that lifetime will depend upon the intensity of the pulse. Experimental studies can reveal the dependence of this lifetime upon pulse parameters such as the intensity, and the center frequency of the pulse.

## IV. CONCLUSION

The model presented here describes population transfer between two crossing electronic levels following the excitation from a chirped pulse. We showed that interference can arise due the wave packet propagating along different pathways in the excited state and their interference. Recent experimental results are now beginning to be able to detect the wave packet branching in the excited state [34]. It was recently shown in experiments on molecular isomerization, using transient absorption methods, that the vibrational frequencies observed had a range of amplitudes. The distribution of vibrational frequency amplitudes could indicate that some modes experience more deconstructive interference than others [82], which also supports this theory.

Chirping the pulse changes the frequency as a function of time and directly affects how the vibrational coherences are initially created in the excited state, which can be represented as Floquet states [47]. Thus the wave packet momentum in the excited state can be changed by chirping the pulse. It is for this reason, we expect that a chirped pulse will be able to detect the interference of various pathways in the potential energy surface.

In the Hanle effect, the application of a magnetic field allows one to measure a lifetime by controlling the interference of the two pathways contributing to the fluorescence. In our case, the interference between wave packets can be controlled through their chirp-dependent momentum. Thus a momentum-dependent lifetime can be obtained. We showed that the lifetime is measureable in the fluorescence signal as a function of chirp. The lifetime can be measured for molecules that show asymmetry between positive vs. negative chirp rates due to the steepness of the potential energy surface [63, 64]. The lifetime should be measured for the positive chirp values only. This is because it is known that positive chirp rates are not sensitive the steepness of the potential energy surface.

The vibrational coherences of the excited state wave packet can be experimentally measured using transient absorption methods [34, 82-87]. Theoretical studies suggest separation of the vibrational coherences into tuning or coupling modes [16]. The identification of the tuning or coupling modes in the experimentally observed transient absorption oscillations is actively being understood. It has been suggested that the tuning modes are more susceptible to wave packet interference. The lower amplitude oscillations can then be identified as tuning modes. The coupling modes are not susceptible to wave packet interference and thus have a higher oscillation amplitude [82]. This suggests that the use of chirped pulses in experiments on molecular isomerization may help to identify tuning and coupling modes. The tuning modes would then show a decrease in amplitude, while increasing the chirp rate. The coupling modes would not show any change with increasing the chirp rate. We hope that the concepts presented here will inspire future experimental work to uncover excited-state molecular processes and wave packet propagation.

## ACKNOWLEDGEMENTS

We would like to thank Benjamin Levine, Yinan Shu, and Artur F. Izmaylov for enlightening discussions on isomerization and conical intersections in molecules. R. Glenn would like to thank M. E. Raikh for discussions about the Hanle effect. This material is based upon work supported by the National Science Foundation under grant CHE-1464807.

# APPENDIX

The geometric phase arises naturally during adiabatic evolution of the system with a cyclic evolution $H(T) = H(0)$, such that after one evolution the eigenstate returns to itself with an associated phase factor $|\Psi(T)\rangle = e^{iC(\gamma)}|\Psi(0)\rangle$ where

$$C(\gamma) = \int_0^T dt \langle \psi(t)|H(t)|\psi(t)\rangle = \gamma_d + \gamma(C) \tag{7}$$

where $\gamma_d$ is a constant and $\gamma(C)$

$$\gamma(C) = \oint_C \langle \psi|i\nabla \psi\rangle dx \tag{8}$$

is the geometric contribution due to propagation around the conical intersection [71]. The effect of the conical intersection, in the Born-Oppenheimer approximation, can be approached by introducing a pseudo magnetic field, i.e., a vector potential into the effective nuclear Hamiltonian, $\nabla \rightarrow \nabla - i(\hat{e}_\phi / 2\rho)$ [71]. The pseudo magnetic field has the form of a magnetic solenoid that is zero everywhere except at the conical intersection (delta-function singularity). The pseudo magnetic field for a solenoid in cylindrical coordinates $(\rho, \phi, z)$ with current flowing along the $z$-axis, leads to a phase $\gamma(C) = \oint_C \frac{1}{2\rho}\rho d\phi = -\pi$, which is why this effect is termed the Molecular Aharonov-Bohm effect [88].

# REFERENCES

[1] G. Moruzzi, and F. Strumia, (Springer Science & Business Media, 2013).
[2] D. R. Yarkony, Rev. Mod. Phys. **68**, 985 (1996).
[3] D. R. Yarkony, J. Phys. Chem. A **105**, 6277 (2001).
[4] B. G. Levine, and T. J. Martínez, Annu. Rev. Phys. Chem. **58**, 613 (2007).
[5] W. Hanle, Z. Physik **30**, (1924).
[6] B. P. Kibble, G. Copley, and L. Krause, Phys. Rev. **153**, 9 (1967).
[7] A. V. Papoyan, M. Auzinsh, and K. Bergmann, Eur. Phys. J. D **21**, 63 (2002).
[8] J. Alnis, K. Blushs, M. Auzinsh, S. Kennedy, N. Shafer-Ray, and E. R. I. Abraham, J. Phys. B: At. Mol. Opt. Phys. **36**, 1161 (2003).
[9] D. R. Crosley, and R. N. Zare, Phys. Rev. Lett. **18**, 942 (1967).
[10] R. S. Ferber, O. A. Shmit, and M. Y. Tamanis, Chem. Phys. Lett. **61**, 441 (1979).
[11] D. Awschalom, D. Loss, and N. Samarth, (Springer Science & Business Media, 2002).
[12] M. I. Dyakonov, (Springer Science & Business Media, 2008).
[13] R. N. Zare, J. Chem. Phys. **45**, 4510 (1966).
[14] L. J. Butler, Annu. Rev. Phys. Chem. **49**, 125 (1998).
[15] J. D. Coe, and T. J. Martínez, J. Am. Chem. Soc. **127**, 4560 (2005).
[16] W. Domcke, D. R. Yarkony, and H. Köppel, (World Scientific, 2011).
[17] Y. Haas, and S. Zilberg, J. Photochem. Photobiol. C **144**, 221 (2001).
[18] D. Sampedro Ruiz, A. Cembran, M. Garavelli, M. Olivucci, and W. Fuß, J. Photochem. Photobiol. **76**, 622 (2002).
[19] M. Nairat, A. Konar, M. Kaniecki, V. V. Lozovoy, and M. Dantus, Phys. Chem. Chem. Phys. **17**, 5872 (2015).
[20] G. Rothenberger, D. K. Negus, and R. M. Hochstrasser, **79**, 5360 (1983).
[21] S. P. Velsko, and G. R. Fleming, Chem. Phys. **65**, 59 (1982).
[22] G. A. Worth, and L. S. Cederbaum, Annu. Rev. Phys. Chem. **55**, 127 (2004).
[23] R. S. Liu, and A. E. Asato, Proc. Natl. Acad. Sci. U.S.A. **82**, 259 (1985).
[24] H. A. Jahn, and E. Teller, Proc. Roy. Soc. A **161**, 220 (1937).
[25] Y. Aharonov, and D. Bohm, Phys. Rev. **115**, 485 (1959).
[26] G. Herzberg, and H. C. Longuet-Higgins, Discuss. Faraday Soc. **35**, 77 (1963).
[27] C. A. Mead, and D. G. Truhlar, J. Chem. Phys. **70**, 2284 (1979).
[28] C. A. Mead, Rev. Mod. Phys. **64**, 51 (1992).
[29] D. R. Yarkony, Acc. Chem. Res. **31**, 511 (1998).
[30] S. Pancharatnam, Proc. Indian Acad. Sci. **44**, 398 (1956).
[31] S. Barry, Phys. Rev. Lett. **51**, 2167 (1983).
[32] M. V. Berry, Proc. R. Soc. Lond. A **392**, 45 (1984).
[33] A. J. Wurzer, T. Wilhelm, J. Piel, and E. Riedle, Chem. Phys. Lett. **299**, 296 (1999).
[34] J. Brazard, L. A. Bizimana, T. Gellen, W. P. Carbery, and D. B. Turner, J. Phys. Chem. Lett. **7**, 14 (2016).
[35] T. Taneichi, T. Kobayashi, Y. Ohtsuki, and Y. Fujimura, Chem. Phys. Lett. **231**, 50 (1994).
[36] J. Schön, and H. Köppel, J. Phys. Chem. **103**, 9292 (1995).
[37] H. Tamura, S. Nanbu, T. Ishida, and H. Nakamura, J. Chem. Phys. **125**, 034307 (2006).
[38] M. Greenfield, S. D. McGrane, and D. S. Moore, J. Phys. Chem. A **113**, 2333 (2009).
[39] G. Vogt, G. Krampert, P. Niklaus, P. Nuernberger, and G. Gerber, Phys. Rev. Lett. **94**, 068305 (2005).

[40] G. Vogt, P. Nuernberger, T. Brixner, and G. Gerber, Chem. Phys. Lett. **433**, 211 (2006).
[41] M. Abe, Y. Ohtsuki, Y. Fujimura, Z. Lan, and W. Domcke, J. Chem. Phys. **124**, 224316 (2006).
[42] D. Geppert, L. Seyfarth, and R. d. Vivie-Riedle, Appl. Phys. B **79**, 987 (2004).
[43] R. d. Vivie-Riedle, L. Kurtz, and A. Hofmann, Pure Appl. Chem. **73**, 525 (2009).
[44] K. Hoki, Y. Ohtsuki, H. Kono, and Y. Fujimura, J. Phys. Chem. A **103**, 6301 (1999).
[45] L. Joubert-Doriol, I. G. Ryabinkin, and A. F. Izmaylov, J. Chem. Phys. **139**, 234103 (2013).
[46] M.-C. Yoon, D. H. Jeong, S. Cho, D. Kim, H. Rhee, and T. Joo, J. Chem. Phys. **118**, 164 (2003).
[47] J. H. Shirley, Phys. Rev. **138**, B979 (1965).
[48] D. Viennot, G. Jolicard, J. P. Killingbeck, and M. Y. Perrin, Phys. Rev. A **71**, 052706 (2005).
[49] D. R. Yarkony, J. Phys. Chem. **100**, 18612 (1996).
[50] G. Karlstrom *et al.*, Comput. Mater. Sci. **28**, 222 (2003).
[51] W. Domcke, D. Yarkony, and H. Köppel, (World Scientific, 2004).
[52] B. G. Levine, C. Ko, J. Quenneville, and T. J. Martinez, Mol. Phys. **104**, 1039 (2006).
[53] M. Thoss, W. H. Miller, and G. Stock, J. Chem. Phys. **112**, 10282 (2000).
[54] X. Sun, and W. H. Miller, J. Chem. Phys. **106**, 916 (1997).
[55] G. Stock, and M. Thoss, Phys. Rev. Lett. **78**, 578 (1997).
[56] N. Minezawa, and M. S. Gordon, **137**, 034116 (2012).
[57] X. Zhang, and J. M. Herbert, J. Phys. Chem. B **118**, 7806 (2014).
[58] A. Kahan, A. Wand, S. Ruhman, S. Zilberg, and Y. Haas, J. Phys. Chem. A **115**, 10854 (2011).
[59] S. Mukamel, (Oxford University Press, 1999).
[60] D. Chruscinski, and A. Jamiolkowski, (Springer Science & Business Media, 2012).
[61] N. V. Vitanov, and S. Stenholm, Phys. Rev. A **55**, 2982 (1997).
[62] S. Stenholm, (Courier Corporation, 2012).
[63] A. Paloviita, Opt. Commun. **119**, 533 (1995).
[64] B. M. Garraway, and K. A. Suominen, Rep. Prog. Phys. **58**, 365 (1995).
[65] G. L. Yudin, A. D. Bandrauk, and P. B. Corkum, Phys. Rev. Lett. **96**, 063002 (2006).
[66] V. S. Malinovsky, and J. L. Krause, Phys. Rev. A **63**, 043415 (2001).
[67] V. S. Malinovsky, and J. L. Krause, Eur. Phys. J. D **14**, 147 (2001).
[68] S. Zhdanovich, E. A. Shapiro, M. Shapiro, J. W. Hepburn, and V. Milner, Phys. Rev. Lett. **100**, 103004 (2008).
[69] Y. B. Band, and Y. Avishai, (Academic Press, 2013).
[70] M. Oppermann, edited by (Springer International Publishing, 2014), pp. 9.
[71] J. W. Zwanziger, M. Koenig, and A. Pines, Annu. Rev. Phys. Chem. **41**, 601 (1990).
[72] K. Datta, Found Phys Lett **2**, 425 (1989).
[73] F. Hasselbach, and M. Nicklaus, Phys. Rev. A **48**, 143 (1993).
[74] J. Du, P. Zou, M. Shi, L. C. Kwek, J.-W. Pan, C. H. Oh, A. Ekert, D. K. L. Oi, and M. Ericsson, Phys. Rev. Lett. **91**, 100403 (2003).
[75] M. Ericsson, D. Achilles, J. T. Barreiro, D. Branning, N. A. Peters, and P. G. Kwiat, Phys. Rev. Lett. **94**, 050401 (2005).
[76] R. Gherib, I. G. Ryabinkin, and A. F. Izmaylov, J. Chem. Theory Comput. **11**, 1375 (2015).
[77] G. A. Voth, and R. M. Hochstrasser, J. Phys. Chem. **100**, 13034 (1996).
[78] C. Zhu, Y. Teranishi, and H. Nakamura, edited by I. Prigogine, and S. A. Rice (John Wiley & Sons, Inc., 2001), pp. 127.
[79] H. Nakamura, (World Scientific, 2012).
[80] H. Tamura, S. Nanbu, H. Nakamura, and T. Ishida, Chem. Phys. Lett. **401**, 487 (2005).
[81] H. Tamura, S. Nanbu, T. Ishida, and H. Nakamura, J. Chem. Phys. **124**, 084313 (2006).
[82] C. Schnedermann, M. Liebel, and P. Kukura, J. Am. Chem. Soc. **137**, 2886 (2015).
[83] S. A. Trushin, T. Yatsuhashi, W. Fuß, and W. E. Schmid, Chem. Phys. Lett. **376**, 282 (2003).
[84] K. A. Kitney-Hayes, A. A. Ferro, V. Tiwari, and D. M. Jonas, J. Chem. Phys. **140**, 124312 (2014).
[85] S. D. McClure, D. B. Turner, P. C. Arpin, T. Mirkovic, and G. D. Scholes, J. Phys. Chem. B **118**, 1296 (2014).
[86] T. A. A. Oliver, and G. R. Fleming, J Phys Chem B **119**, 11428 (2015).
[87] P. C. Arpin *et al.*, J. Phys. Chem. B **119**, 10025 (2015).
[88] C. Alden Mead, Chem. Phys. **49**, 23 (1980).